\documentclass[structabstract]{aa}
\usepackage{txfonts}
\usepackage{natbib}
\usepackage{graphicx}

\newcommand{\OII}{[O{\tt II}]$\lambda$3727}
\newcommand{\oii}{[O{\tt II}]}
\newcommand{\OIII}{[O{\tt III}]$\lambda$5007}
\newcommand{\NII}{[N{\tt II}]$\lambda$6584}

\begin{document}


\title{
The zCOSMOS redshift survey: the three-dimensional classification cube
     and bimodality in galaxy physical properties 
   \thanks{Based on observations undertaken at the European
   Southern Observatory (ESO) Very Large Telescope (VLT) under Large
   Program 175.A-0839.  Also based on observations with the NASA/ESA
   Hubble Space Telescope, obtained at the Space Telescope Science
   Institute, operated by AURA Inc., under NASA contract NAS 5-26555,
   with the Subaru Telescope, operated by the National Astronomical
   Observatory of Japan, with the telescopes of the National Optical
   Astronomy Observatory, operated by the Association of Universities
   for Research in Astronomy, Inc. (AURA) under cooperative agreement
   with the National Science Foundation, and with the Canada-France-Hawaii
   Telescope, operated by the National Research Council of Canada,
   the Centre National de la Recherche Scientifique de France
   and the University of Hawaii.}
}

\author{
M.~Mignoli\inst{1}
\and
G.~Zamorani\inst{1}
\and 
M.~Scodeggio\inst{2}
\and
A.~Cimatti\inst{3}
\and
C.~Halliday\inst{4}
\thanks{\emph{Present address:} Department of Physics and Astronomy, University of Glasgow, 
Glasgow G12 8QQ, United Kingdom}
\and
S.J.~Lilly\inst{5}
\and
L.~Pozzetti\inst{1}
\and
D.~Vergani\inst{1,2}
\and
C.~M.~Carollo\inst{5}
\and
T.~Contini\inst{6}
\and
O.~Le~F\'evre\inst{7}
\and
V.~Mainieri\inst{8}
\and
A.~Renzini\inst{9}
\and
S.~Bardelli\inst{1}
\and
M.~Bolzonella\inst{1}
\and
A.~Bongiorno\inst{10}
\and
K.~Caputi\inst{5}
\and
G.~Coppa\inst{3,1}
\and
O.~Cucciati\inst{11}
\and
S.~de~la~Torre\inst{7}
\and
L.~de~Ravel\inst{7}
\and
P.~Franzetti\inst{2}
\and
B.~Garilli\inst{2}
\and
A.~Iovino\inst{11}
\and
P.~Kampczyk\inst{5}
\and
J.-P.~Kneib\inst{7}
\and
C.~Knobel\inst{5}
\and
K.~Kova\v{c}\inst{5}
\and
F.~Lamareille\inst{6,1}
\and
J.-F.~Le~Borgne\inst{6}
\and
V.~Le~Brun\inst{7}
\and
C.~Maier\inst{5}
\and
R.~Pell\`o\inst{6}
\and
Y.~Peng\inst{5}
\and
E.~Perez~Montero\inst{6}
\and
E.~Ricciardelli\inst{9}
\and
C.~Scarlata\inst{5}
\and
J.D.~Silverman\inst{5,8}
\and
M.~Tanaka\inst{8}
\and
L.~Tasca\inst{7}
\and
L.~Tresse\inst{7}
\and
E.~Zucca\inst{1}
\and
U.~Abbas\inst{7}
\and
D.~Bottini\inst{2}
\and
P.~Capak\inst{12}
\and
A.~Cappi\inst{1}
\and
P.~Cassata\inst{7}
\and
M.~Fumana\inst{2}
\and
L.~Guzzo\inst{11}
\and
A.~Leauthaud\inst{7}
\and
D.~Maccagni\inst{2}
\and
C.~Marinoni\inst{13}
\and
H.J.~McCracken\inst{14}
\and
P.~Memeo\inst{2}
\and
B.~Meneux\inst{11}
\and
P.~Oesch\inst{5}
\and
C.~Porciani\inst{5}
\and
R.~Scaramella\inst{15}
\and
N.~Scoville\inst{12}
}

\institute{
INAF -- Osservatorio Astronomico di Bologna, Bologna, Italy
\and
INAF -- Istituto di Astrofisica Spaziale e Fisica Cosmica, Milano, Italy
\and
Dipartimento di Astronomia, Universit\`a degli Studi di Bologna, Bologna, Italy
\and
INAF -- Osservatorio Astronomico di Arcetri, Firenze, Italy
\and
Institute of Astronomy, ETH Zurich, Zurich, Switzerland
\and
Laboratoire d'Astrophysique de Toulouse-Tarbes, Universit\'e de Toulouse,
CNRS Toulouse, France, %14 av.~Edouard Belin, F-31400 
\and
Laboratoire d'Astrophysique de Marseille, Marseille, France
\and
European Southern Observatory, Garching, Germany
\and
Dipartimento di Astronomia, Universit\`a di Padova, Padova, Italy
\and
Max Planck Institut f\"ur Extraterrestrische Physik, Garching, Germany
\and
INAF -- Osservatorio Astronomico di Brera, Milano, Italy
\and
California Institute of Technology, Pasadena, CA, USA
\and
Centre de Physique Theorique, Marseille, Marseille, France
\and
Institut d'Astrophysique de Paris, Universit\'e Pierre \& Marie Curie, Paris, France
\and
INAF -- Osservatorio Astronomico di Roma, Monte Porzio Catone , Italy
}

\date{Received  / Accepted }

\abstract {} 
{We investigate the relationships between three main optical
galaxy observables (spectral properties, colours, and morphology),
exploiting the data set provided by the COSMOS/zCOSMOS survey.
The purpose of this paper is to define a simple galaxy classification cube,
using a carefully selected sample of $\approx1000$ galaxies.} 
{Using medium resolution spectra of the first
$1k$~zCOSMOS-bright sample, optical photometry from the
Subaru/COSMOS observations, and morphological measurements derived
from ACS imaging, we analyze the properties of the galaxy
population out to $z\sim 1$. Applying three straightforward classification
schemes (spectral, photometric, and morphological),
we identify two main galaxy types, which appear to be
linked to the bimodality of galaxy population. The three parametric
classifications constitute the axes of a ``classification cube''.} 
{A very good agreement exists between the classification from
spectral data (quiescent/star-forming galaxies) and that based on colours
(red/blue galaxies). The third parameter (morphology) is less well
correlated with the first two: in fact a good correlation between the
spectral classification and that based on morphological analysis
(early-/late-type galaxies) is achieved only after partially complementing
the morphological classification with additional colour information.
Finally, analyzing the 3D-distribution of all galaxies in the sample,
we find that about 85\% of the galaxies show a fully concordant
classification, being either quiescent, red, bulge-dominated galaxies
($\sim$~20\%) or star-forming, blue, disk-dominated galaxies ($\sim$~65\%).
These results imply that the galaxy bimodality is a
consistent behaviour both in morphology, colour and dominant stellar
population, at least out to $z\sim 1$.}
{}

\keywords{galaxies: general -- galaxies: evolution -- galaxies:
fundamental parameters (classification)}

\titlerunning{Galaxy Bimodality in 3D}
\authorrunning{Mignoli et al.}

\maketitle
%
% The main text
%
\section{Introduction}

Galaxies have a wide variety of physical and observational
properties. However, since the early works of \citet{hum36} and
\citet{hub38} it has been known that the morphologies of galaxies
correlate with their colours and consequently with the derived
properties of their stellar population.
Large efforts have been undertaken to establish 
reliable relationships between the morphological types, colours, and
spectral characteristics of the galaxies \citep[i.e.][]{fio99,ber00}. 
The accurate determination of these relationships is of paramount
importance to our understanding of galaxy formation and evolution.
The main galaxy optical properties are the spectral energy
distributions, spectral line properties, internal velocity dispersions,
morphologies, sizes, and structural components. From these observable
quantities, we can derive intrinsic physical properties,
such as stellar and gaseous masses, metallicities, and star-formation
histories; each of these aspects provides crucial clues
about how galaxies were created and have evolved.
The most basic evolutionary processes and their corresponding observables
are, however, not yet clear and we are still uncertain whether
the correlations between galaxy properties observed are driven
by fundamental aspects of the galaxy population \citep{cons06}.
The goal of this paper is to investigate the relationships between the
three main optical galaxy observables (spectral properties, colours,
and morphology) by taking advantage of the unique
data set provided by the COSMOS/zCOSMOS survey.

A fundamental property discovered by large surveys such
as the Sloan Digital Sky Survey (SDSS) and the Two-Degree Field
Galaxy Redshift Survey (2dFGRS), is that the optical colour
distribution of galaxies at low redshift can be represented by a bimodal
function \citep{stra01,wild05}. The bimodality separates the
locus of star-forming galaxies from that of early-type galaxies
(ellipticals, $S0$, and early-type spirals), such that the fraction of
the red, passive population increases progressively with the
luminosity, becoming smaller than the blue fraction for $M_r>-20$ and
dominating the distribution for $M_r<-21$ \citep{bald04}. 
The bimodality is also present in different aspects of the galaxy
population: structural parameters such as concentration index 
\citep{ball06} and disk-to-bulge ratio \citep{dro07},
and spectral features such as $D4000$ and line ratios 
\citep{kau03,bal04a,yan06,fra07}, all exhibit strong bimodal distributions. 
The bimodality in galaxy properties, clearly evident in the local Universe,
is still present at $z\sim1$ \citep{bell04}, but it is less clear whether
the two distinct galaxy populations %, with really distinct properties,
hold at higher redshifts \citep[up to $z=2$;][]{giallo05,cir07,cassa08}.
A key test for galaxy evolution models is to explain why this bimodality
occurs; it is still unclear which processes produce two sets of galaxies
of different average colours and dispersions and different luminosity ranges.

Determining the  nature of the bimodality can be approached
by means of large spectroscopic redshift surveys, studying
the measured galaxy parameters as a function of redshift, luminosity,
and spectral properties. The zCOSMOS survey opens the possibility
to extend the study of bimodality out to $z\sim1$.

zCOSMOS is a large redshift survey that is being undertaken in the
COSMOS field as an ESO Large Programme ($\sim600$ hours of observation)
with the VIMOS spectrograph at the VLT. The survey consists of two
parts: the first part targets a magnitude-limited sample selected from
ACS images to have $I_{AB} < 22.5$ and observed at medium resolution;
this will provide spectra and redshifts for $\approx20,000$ galaxies
out to $z\sim1.5$ over two square degrees.
The second and deeper part of the survey is acquiring
spectroscopy to measure the redshifts of
$\approx12,000$ galaxies at $1.4<z<2.5$ at lower resolution and
over the central square degree, by selecting targets 
using various colour-selection criteria.

This paper, based mainly on the spectroscopic data of the
zCOSMOS survey, takes advantage of the large potential of the COSMOS database.
We present the analysis of a representative sample
of the bright component of the full survey, consisting of around
1200 galaxies observed during the first year of observations.
Using the medium resolution spectra of these objects and
complementing them with the optical photometry
from the COSMOS multi-band catalog \citep{cap07} and morphological
information derived from ACS imaging \citep{scov07,sca07a},
we investigate the optical properties of the galaxy population
out to $z\sim 1$. Applying three straightforward classifications
(spectral, photometric, and morphological), we are able to distinguish
two main galaxy types, which are intimately connected with the
bimodality and formation history of the galaxy population: the three
parametric classifications constitute the axes of a ``classification cube''.
It is important to remark that we develop our cube using only
observational parameters, such as emission-line equivalent widths,
continuum indices, photometric colours (and not synthetic colours), 
and a non-parametric morphological classification. The choice
is motivated by the desire to avoid dependence on models,
galaxy templates, and parametrization of galaxy properties.
Moreover, using simple observed properties makes the classification
cube easy to replicate in other galaxy surveys.

The principal aim of this paper is the definition of the galaxy
classification cube, using a fair but relatively small sample of
$\approx1000$ galaxies.  The full exploitation of the classification
cube, applied to larger galaxy samples (i.e. the 10K zCOSMOS sample,
Lilly et~al. 2008, in preparation), will be presented in a
forthcoming paper (Coppa et~al., in preparation).

\section{Description of zCOSMOS observations and data reduction}

For an exhaustive discussion of the zCOSMOS survey design, in particular
of the assembly of the input catalogues, the target selection, and 
the spectroscopic mask design, we address the reader to the survey
presentation paper \citep{lil07}.  

%The zCOSMOS observations are executed at the VLT in Service Mode. 
The brighter part of zCOSMOS project consists of spectroscopic data
of targets identified by means of a pure magnitude-limited selection
$I_{AB} < 22.5$, as used in the CFRS \citep{lil95} 
and VVDS-wide surveys \citep{olf05}. This selection culls
galaxies mainly in the redshift range $0.1 < z <1.2$.
A medium resolution grism ($R\sim 600$) has been used
with a slit width of 1~arcsec, %in order
to achieve a velocity accuracy of $\sim$100 km$s^{-1}$, and with
one hour integrations to enable redshifts to be measured with a
high success rate. 
The spectral range of observations is typically $5550 - 9650$~\AA,
which enables important spectroscopic diagnostic lines, centered on
a rest-frame wavelength of approximately 4000\AA, to be measured
out to a redshift of about 1.
The one hour total integration is acquired in five $720\,s$ exposures,
between which the telescope was placed at different offset positions
to complete an optimal sky subtraction, 
in a pattern of positions separated by one arcsec along the slit.
%in order to optmize the sky background removal.
The observations were executed during periods when the seeing was
better than 1.2 arcsec.
The first zCOSMOS-bright spectroscopic observations were completed
in VLT Service Mode between April and June 2005. Eight sets of VIMOS
mask observations were completed, which provided spectra of 1303 objects,
corresponding to about 6\% of the planned bright survey program.
The data reduction was carried out using the {\it VIPGI} software,
developed specifically for data acquired by VIMOS instrument \citep{sco05}.
\citet{lil07} described the redshift determination procedures
and the assignment of confidence classes to each measured redshift.
The confidence in the redshift determination
was indicated by a flag, whose values were: 4
(completely secure redshift); 3 (very secure redshift, but
with a very marginal possibility of error); 2 (a likely
redshift, but with a significant possibility of error); 1
(possible redshift); 0 (no redshift determination); and 9
(redshift based on a single strong emission line).
The high quality of the photometric measurements for the COSMOS field
and the corresponding photometric redshifts at $z < 1.2$ \citep{fel06},
implied that the photometric redshifts could be used to assess the
spectroscopic redshift reliability of our zCOSMOS-bright spectra. 
To indicate the reliability of a redshift, we added a decimal place
to its confidence flag: if the difference between the photometric and
spectroscopic redshift, $|z_p - z_s|$, was less than $0.1\times(1+z_s)$,
we added the decimal place=.5; otherwise, we added a decimal place=.1.
In the following, we include in the spectroscopic ``high quality''
data set all the galaxy spectra with flag 3 and 4, and those with flag
2.5, which includes galaxy spectra with flag=2 confidence for which
the photometric and spectroscopic redshifts are in good agreement.
We exclude from the high quality data set galaxy spectra of
flag 9, where the redshift could be determined using only one
emission line; although this flag corresponds to significant
detection of an emission line, the absence of other spectroscopic
diagnostic lines implies that for many the continuum signal-to-noise
ratio (hereafter S/N) could be rather low.
%The possible values for the decimal places are: 5 (the photometric
%redshift is consistent with the spectroscopic redshift within
%|zphot - zspec| < 0.10 � [1 + zspec]); 4 (no zphot avail-
%able; usually for galactic stars or quasi-stellar objects);
%3 (only for flag=9 objects; indicates that zphot solves
%the \oii - H$\alpha$ degeneracy); and 1 (the zphot
%is inconsistent with the zspec at a 0.10�[1+zspec] level).
%Hereinafter, we claim consistency between photometric and
%spectroscopic redshifts if the redshift difference $|zp - zs|$
%is less than $0.1\times(1+zs)$.

\section{Spectral Analysis}

\subsection{Spectral Measurements} \label{spemea}

We performed spectral measurements for a total of 1130 galaxies
(and 20 QSOs) observed during the first year of observations and
for which a redshift had been determined. This set of
extragalactic objects forms the analyzed $1k$~zCOSMOS-bright sample. 

%
% Figure 1
%
\begin{figure}[t]
\resizebox{\hsize}{!}{\includegraphics{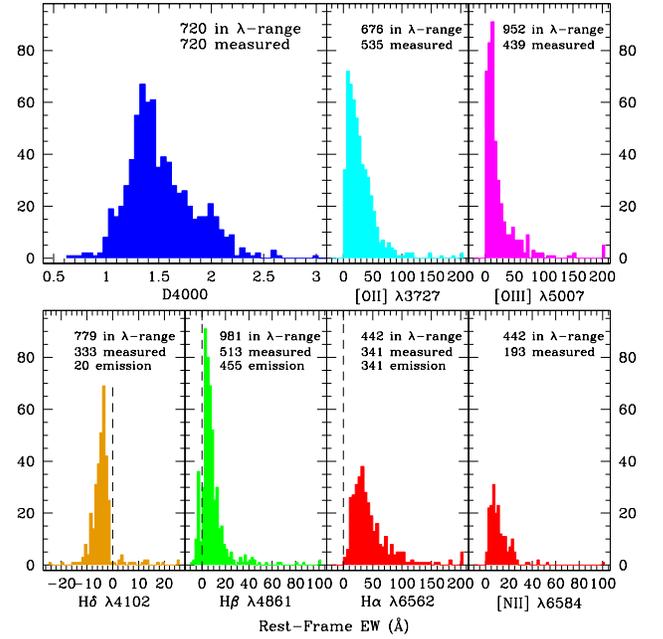}}
\caption{Rest-frame measurement distributions of the analyzed spectral
features, from top-left to bottom-right: $D4000$, and EW of the lines 
\OII, \OIII, H$\delta$, H$\beta$, H$\alpha$, and \NII. 
In each panel, we indicate the number of galaxies for which the feature 
is within the spectral observed range and the number of measured lines. 
\label{figew0}}
\end{figure}

We measured the spectral quantities using semi-automatic procedures 
that were adaptations of the IRAF task {\tt splot}.
First, the continuum was fitted automatically to fixed wavelength
intervals, although our procedures also enabled interactive adjustments
of the continuum level to improve the line measurement in
noisy spectra. Then, equivalent widths (EWs) and fluxes were
measured by both applying a Gaussian-function fitting algorithm
and direct integration of the continuum-subtracted line profiles.
The fluxes measured by both techniques were in excellent agreement:
for all emission lines, the two measurements agreed to within 25\%,
while the 1-$\sigma$ dispersion of the difference distribution 
corresponded to only 11\%. Figure~\ref{figew0} shows the distributions
for principal spectroscopic line measurements for the $1k$-sample.
The equivalent widths are given in the rest-frame, and indicated
by positive or negative values for emission or absorption lines,
respectively. Errors in the measured equivalent widths 
were estimated by taking into account both the measured
r.m.s. of the continuum close to the lines and the difference
between the values derived by the two techniques. 
The median uncertainty in an individual EW is 0.7~\AA, but it is
worth noting that we did not account for the stellar absorption
in the measurements of the Balmer emission lines, since in most
cases the S/N of our spectra did not enable a simultaneous
fit of both the absorption and emission components.
We measured EWs and fluxes for \OII, H$\delta$, H$\beta$,
\OIII, H$\alpha$, and \NII, as well as the strength of the
4000~\AA\ break \citep[$D4000$;][]{bru83}.
The continuum fluxes were measured in fixed wavelength ranges,
and our procedure computed the average using a sigma-clipping
technique, to ensure that spikes, due to bad sky subtraction and/or to 
cosmic-ray residuals, did not affect significantly the measured values.
Errors in the continuum fluxes were then computed from the standard
deviation. 
%We focused our analysis on EW measurements and flux ratios,
%since the absolute spectrophotometry is not completely reliable because
%the spectra were not corrected for slit losses.
We focused our analysis on EW measurements and flux ratios 
because the absolute spectrophotometry was not completely reliable 
due to the corrections for slit losses that had been applied to the spectra.
This set of ``semi-automatic'' measurements was used, in addition,
to test and calibrate the fully automatic measurement code PlateFit
\citep{lam06}, which is being applied to larger zCOSMOS data sets.

%
% Figure 2
%
\begin{figure}[ht]
\resizebox{\hsize}{!}{\includegraphics{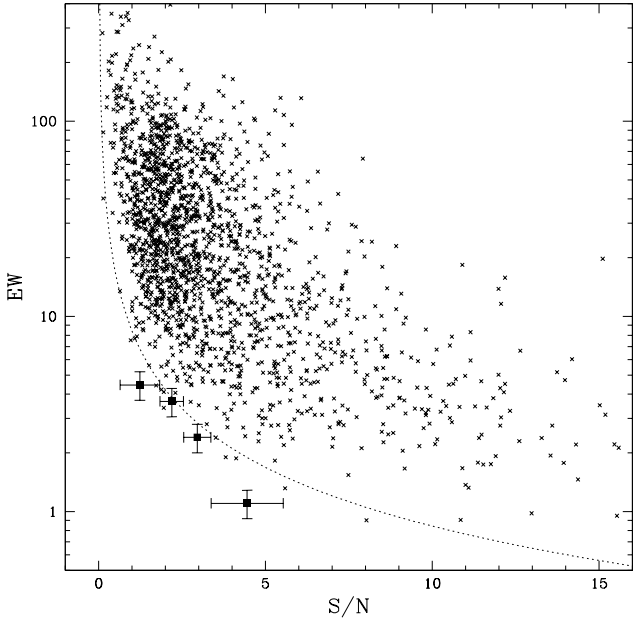}}
\caption{Observed EWs of all the measured emission lines
(see Fig.~\ref{figew0}) versus the signal-to-noise ratio
of the continuum close to the line; the lower envelope of the measured
equivalent widths is fitted well by the dotted curve
(see Eq.~\ref{eqewup} in the text). The larger filled square symbols
represent the \oii \ EW values measured in composite spectra of galaxies
that individually did not demonstrate a detectable emission line,
grouped in four equally populated bins according to their continuum
S/N; the line EWs have been measured in the composite spectra shifted
to mean redshift of the corresponding galaxy group.
The horizontal error bars indicate the range of S/N included
in each bin, whereas the vertical error bars show the variation
in the  observed EW if the composite
spectra were moved to the extremes of the redshift range. 
\label{figew_upper}}
\end{figure}

In order to estimate our emission line detection limits, we measured
the signal-to-noise ratio in the continuum adjacent to the lines.
Figure~\ref{figew_upper} shows the relation between the S/N in
the continuum and the observed equivalent widths of the measured
emission lines in our sample: from this figure, it is clear that a
correlation exists between the minimum detected EW and the continuum S/N.
%the lines have been detected down to a observed equivalent
%width of $\approx 1$~\AA \ for the bulk of the zCOSMOS-bright galaxies.
The lower envelope of the measured EWs (dotted line in
Fig.~\ref{figew_upper}) is well described by the curve

\begin{equation}
EW(detection\;limit) = {SL*\Delta\over{(S/N)_{cont}}}
\label{eqewup}
\end{equation}

\noindent
where $\Delta$ is the resolution element (in \AA) of our spectra
and $SL$ is the significance level %\footnote{The significance level
%is the probability of incorrectly rejecting a false hypothesis.}
of the detectable line, expressed in terms of sigma of the continuum
noise \citep{LDSS,man02}.
On the basis of Fig.~\ref{figew_upper}, we adopt a $SL(\sigma)=3.5$.
Measuring the S/N in the continuum of each galaxy spectra without
a detectable emission line, we estimate the EW upper limits 
(using Eq.~\ref{eqewup}). The robustness
of this technique in estimating the line upper limits is 
demonstrated by the detection of the emission lines (in particular
\OII) in composite spectra of galaxies without a detectable
emission line. We grouped 141 single spectra in four equally populated
bins according to their continuum S/N, and all the measured \oii \ EWs 
lie below the curve calculated using the above equation (large
squares in Fig.~\ref{figew_upper}).

\subsection{Galaxy Spectral Classification} \label{specls}

%
% Figure 3
%
\begin{figure}[ht]
\resizebox{\hsize}{!}{\includegraphics{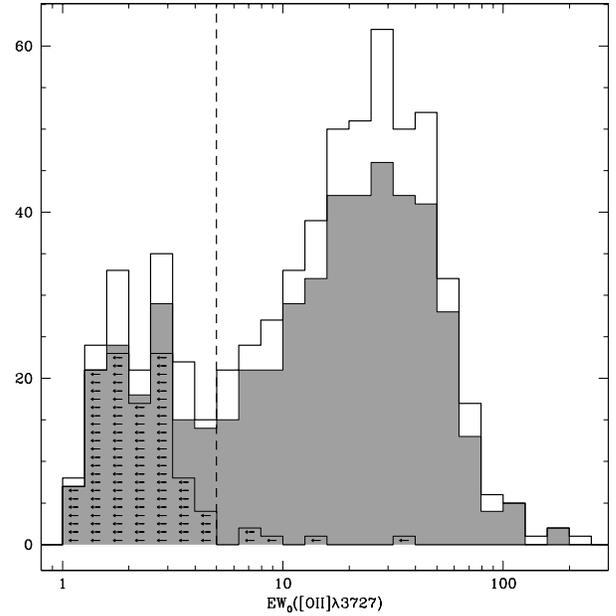}}
\caption{Distribution of the rest-frame \oii \ EWs  %(upper panel)
for the zCOSMOS galaxy $1k$-sample. 
The gray histogram represents the parameter distribution for
galaxies included in the high-quality data set, with leftward
arrows indicating the upper limits calculated using
Eq.~\ref{eqewup}; the thin empty histogram is for the whole sample. 
\label{fig_bimo}}
\end{figure}
%
% Figure 4
%
\begin{figure*}[p]
\centering
\resizebox{0.67\hsize}{!}{\includegraphics{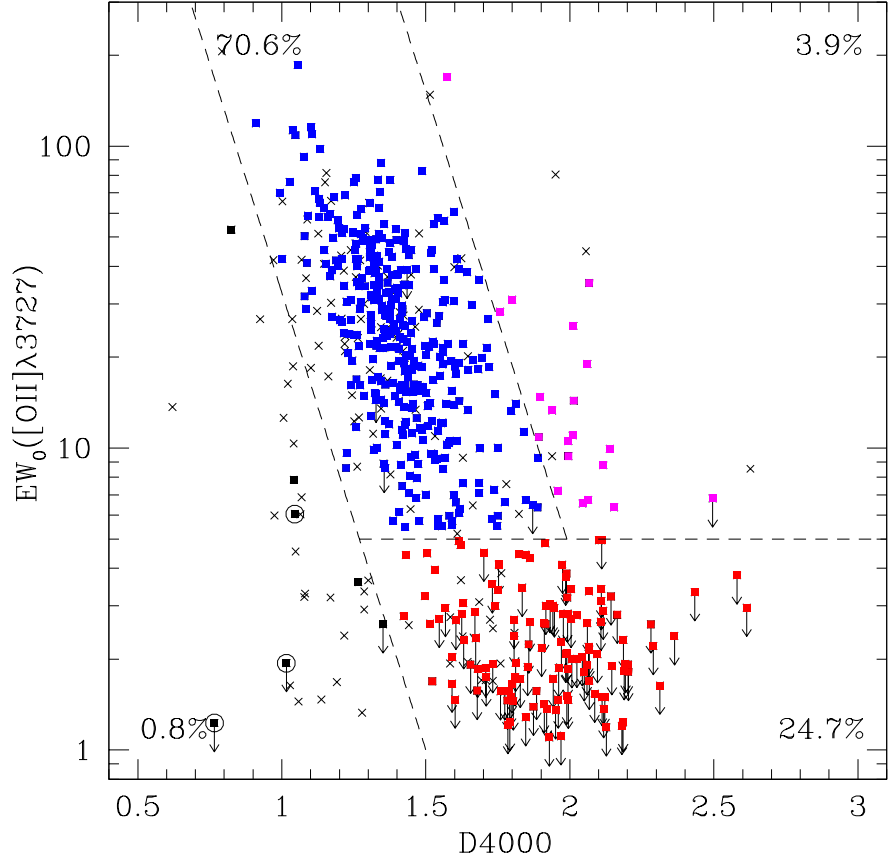}}
\caption{The classification plane for the 1k~zCOSMOS galaxies;
using the rest-frame \OII \ EW and $D4000$, we define three
main classes: {\it red quiescent galaxies} (lower right region), 
{\it blue emission-line galaxies} (upper left region) and 
the {\it intermediate galaxies}, which show emission lines but
a large value of $D4000$. In the figure, we plot all classifiable
galaxies (with $z\sim$~0.45$\div$1.25), with different 
symbols following the redshift quality ranks (crosses for the
low-quality spectra, filled squares for the high-quality ones).
Broad line objects (quasars) are encircled in figure. 
The dashed lines represent the boundaries of the adopted
classification scheme following Eq.~\ref{eqspcls}.
The percentage of high quality galaxies falling in each 
classification regions are labelled at plot corners. 
\label{fig_speplane}}
\end{figure*}
%
% Figure 5
%
\begin{figure*}[p]
\centering
\resizebox{0.67\hsize}{0.45\hsize}{\includegraphics{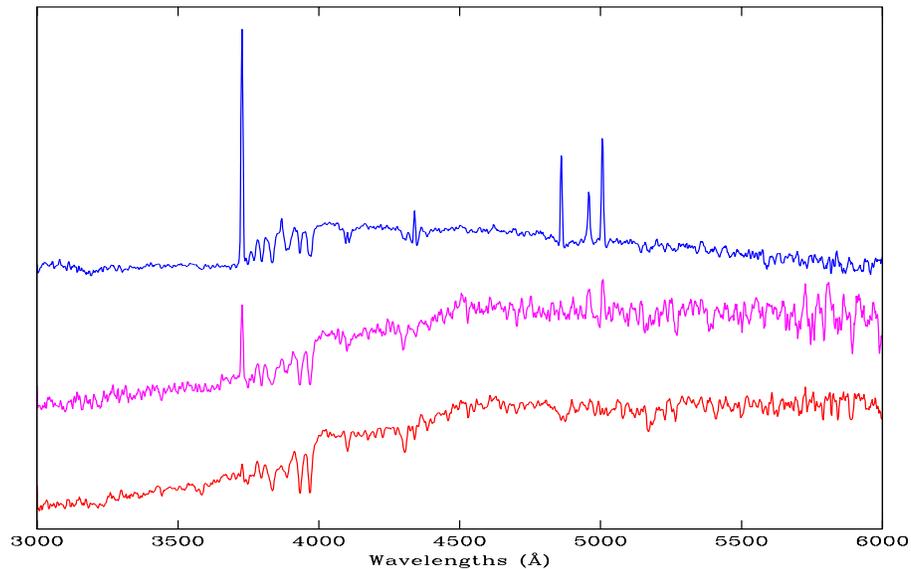}}
\caption{Composite spectra of the three main galaxy classes as defined 
in Fig.~\ref{fig_speplane}. From bottom to top: 126 red quiescent,
20 intermediate and 361 blue star-forming galaxies contribute to the
averages.
\label{fig_compo}}
\end{figure*}

Following \citet{mig05}, we applied to the zCOSMOS-bright galaxy data set
the straightforward spectral classification scheme adopted
for the $K20$ survey \citep{cim02}, which uses the equivalent width of
the \OII \ line and the 4000~\AA \ break index to define two main
spectral classes: {\it red quiescent galaxies}, which show a
large $D4000$ value and a faint, if not undetected, \oii \ emission line;
and {\it blue star-forming galaxies},
with small $D4000$ and intense emission lines. 
This spectral classification scheme naturally arises from the intrinsic
bimodal nature of galaxy physical properties that has been observed
in the colour distribution \citep{bal04b,bell04} and
mean spectroscopic indices \citep{kau03,fra07}.
In Fig.~\ref{fig_bimo}, the distribution of the rest-frame \oii \
equivalent widths is shown for the zCOSMOS galaxy $1k$-sample.
A clear bimodality is evident in the EW$_0$(\oii) distribution,
especially if we consider that the histogram with leftward
arrows represents the upper limits to line measurements.
On the basis of this behaviour,
we separate the galaxy population into two classes, strong and weak
line emitters, divided by an EW$_0$(\oii)=5\AA. With this
threshold, almost all the objects with undetected emission lines are
included in the weak emission line galaxy group. Most galaxies
with high EW that are not included in the high-quality sample
are flag=9 objects (see end of Sect.~2).

Looking at Fig.~\ref{fig_speplane}, a couple of comments can be made.
First, the ``low-density'' region in the leftward corner of
the classification plane is populated mainly by poor-quality
flagged objects (indicated by crosses in the figure), confirming
both the doubtful nature of their redshift identification and the
efficiency of the spectral classification scheme. The galaxies in this
region of the plane are not spectroscopically classified. Furthermore,
when applied to the high-ranked zCOSMOS galaxies, the classification
scheme demonstrates its efficiency: only four objects fall marginally
in the leftward region of the plot, whereas the intermediate
class (red emission line galaxies) constitutes only 4\% of the sample.
Interestingly, the few (broad line) active nuclei observed
in the useful redshift range are quite segregated in the
classification plane (see the encircled symbols in
Fig.~\ref{fig_speplane}), having very blue continuum but
relatively faint \oii \ emission lines \citep[$<$10\AA,][]{for01}.

Figure~\ref{fig_speplane} shows the resulting spectral classification
plane. Due to the high quality of zCOSMOS spectroscopic
data, higher than in the $K20$ survey, we are able to appreciate
how most of the galaxies with intense emission lines
occupy a narrow region in the $D4000$-\oii \ plane.
We performed an iterative linear least squares fit with %2.7-
$\sigma$-clipping, to define and delimitate the locus of these
objects (see diagonal dashed lines in Fig.~\ref{fig_speplane}).
The final criterion adopted to define the blue emission line galaxies was

\begin{eqnarray}
EW_0([\rm O{\tt II}]) \ge 5{\rm \AA}, \nonumber\\  
1.50 \le D4000+0.33\times\log\left(EW_0([{\rm O{\tt II}]})\right) \le 2.22
\label{eqspcls}
\end{eqnarray}

A relationship between the $D4000$ and the \oii \ equivalent width
was expected, since both features are related to the
galaxy star-formation history, either the most recent one
($EW_0([\rm O{\tt II}]$)) or its integral over time ($D4000$).
It is particularly encouraging that, by projecting
our linear fit in the $D4000$-specific SFR plane, with the simple
assumption of constant mass-to-luminosity ratio, we reproduce
almost exactly the mode of the SDSS data distribution presented
in Fig.~11 of \citet{bri04}. 

It is worth keeping in mind that the observed spectral range enables us
to apply this spectral classification scheme only to galaxies of
sufficiently high redshift to include the \oii--$D4000$ region,
i.e. the redshift range $z\sim 0.45 - 1.25$. 
Within this redshift range, 631 galaxies
were classified, of which 511 belong to the high-quality sample
on the basis of the assigned redshift flag.
More than 80\% of the spectroscopically classified zCOSMOS galaxies
belong to the two previously defined classes: {\it red quiescent
galaxies} and {\it blue star-forming galaxies}.
A third group of objects, which have emission lines but a red stellar
continuum (as implied by their high value of $D4000$), constitutes
what we refer to as the intermediate class of galaxies. Interestingly,
20\% of these objects show some sign of nuclear activity: three of the
twenty galaxies in this class are classified as Seyfert~2 galaxies
(Bongiorno et~al., in preparation)  on the basis of the diagnostic diagrams
of \citet{bald81}, and one additional galaxy has an X-ray counterpart
\citep{brusa07}. 
In contrast, the fraction of the blue star-forming galaxy population that
are candidate AGN is only 2\%, i.e. 7 out of 361 galaxies. 
%Comparatively, the same fraction of putative
%AGN is only 2\% (7/361) of the blue starforming galaxy population.

%
% Figure 6
%
\begin{figure*}[t]
\centering
\resizebox{0.67\hsize}{!}{\includegraphics{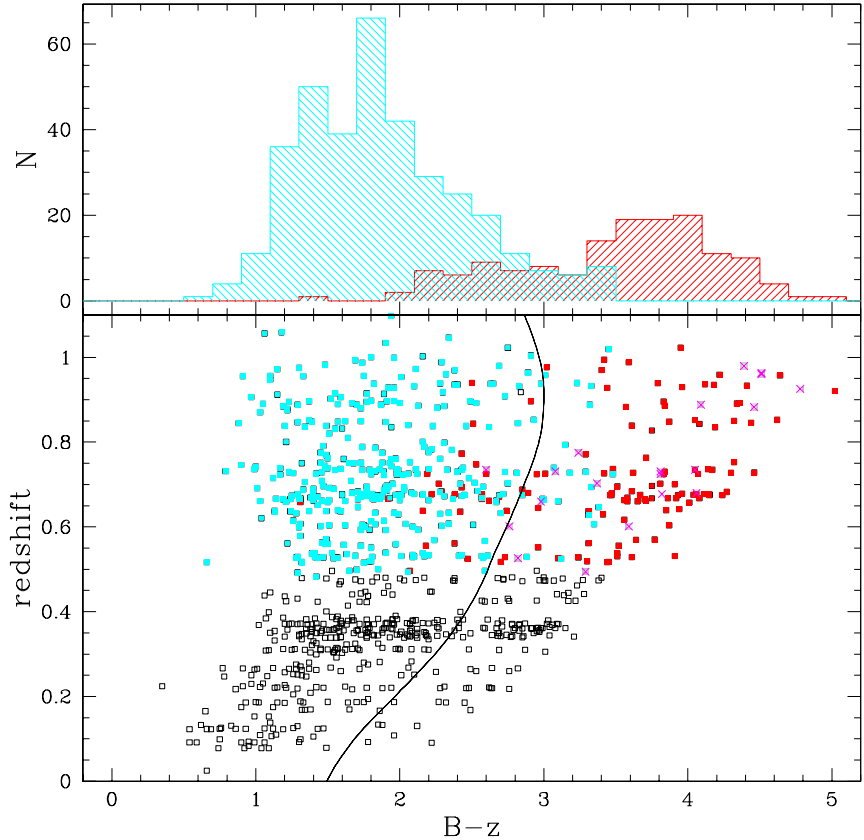}}
\caption{Upper Panel: observed $B-z$ colour distribution of the galaxies
belonging to zCOSMOS-bright $1k$-sample with \hbox{$0.45\le z \le1.2$};
cyan histogram for blue star-forming, red histogram for quiescent
galaxies, both with and without emission lines. Lower Panel: $B-z$ versus
redshift plot for the whole high quality sample; cyan and red squares
for star-forming and quiescent galaxies respectively, magenta
crosses for intermediate ones). 
We divide the two galaxy populations using as discriminator the $B-z$ 
track of the Sab template from the Coleman Wu \& Weedman Extended set
(solid line in figure; \citet{saw97}).
\label{fig_Bz_sep}}
\end{figure*}

Composite spectra for each galaxy class were generated by
averaging all spectra included in that class. 
To create the composites, each spectrum was 
shifted to the rest-frame according to its redshift
(with a 1.5~\AA \ rest-frame bin\footnote{at the median redshift of
the analyzed galaxy sample ($z=0.724$), a rest-frame bin of 1.5~\AA \ 
matches the pixel size used in the spectroscopic
observations.})  %%%%%%%%%%%pixel size = 2.553
and normalized in the 4000-4500~\AA \ wavelength range, which was
always present in the observed spectroscopic window. 
An identical weight was assigned to each individual spectrum, 
to avoid biasing the final composite towards the brightest galaxies.
Finally, all spectra belonging to a particular spectroscopic class
were stacked. In Fig.~\ref{fig_compo}, the average spectra of
the three main spectroscopic classes are plotted.
The continuum in the composite of the emission line galaxies with red
$D4000$ is largely indistinguishable from that of the purely passive
galaxies composite. Therefore, this class of intermediate objects
should consist mainly of galaxies undergoing a modest
star-formation episode, rather than of heavily reddened star-forming
galaxies. The presence of the [OIII] emission doublet,
with no evidence of $H\beta$ in emission, in the composite of
intermediate objects could indicate some nuclear activity
(i.e. type~II AGNs) contamination in this galaxy class, as
already noted above \citep[see also][]{yan06}.

\section{Photometric colours of galaxy spectral classes}

The bimodality in the galaxy spectroscopic index measurements
identified above may be naturally linked to the colour bimodality
observed widely in the galaxy population.
We therefore compare the observed colours and spectral classification
of galaxies that belong to the zCOSMOS-bright $1k$-sample.

The photometric observations of the COSMOS field 
that we use in our study include optical ($B_J$, $g^+$, $V_J$,
$r^+$, $i^+$, and $z^+$ data acquired using SuprimeCam at the
Subaru Telescope, $u^*$-band data with the Canada--France--Hawaii Telescope
and $i_{814}$ images with ACS/HST) and near-Infrared
($K_s$-band data acquired using the Cerro Tololo International Observatory
and Kitt Peak National Observatory telescopes) wavelengths. 
Details of the ground-based observations and data
reduction are presented in \citet{cap07} and \citet{tan07}.
The reduced images in all bands were PSF-matched by Gaussian
convolution with a final FWHM corresponding to the seeing
in the $K_{s}$-band (1.5\arcsec); the multi-band photometry catalog
was then generated using SExtractor \citep{sex96}. 
First, the total (SExtractor magauto) and fixed-aperture
(3\arcsec--diameter) magnitudes were measured on the detection
image (i-band), allowing the estimate of the aperture correction
for each galaxy.
This correction was applied subsequently to aperture magnitudes 
measured in the other optical/near-IR bands. Details of the
photometry, star/galaxy separation, and catalog construction
were provided by \citet{cap07}. 

We analyzed different colours from the photometric catalog and
attempted to identify a separation in photometric space between
the above defined spectral galaxy classes. The $B-z$ colour,
in addition to both $B-K$ and $u-i$, appears to provide the most
effective way of separating red quiescent and blue star-forming galaxies. 
In Fig.~\ref{fig_Bz_sep}, we illustrate, in the $B-z$ distribution
histograms (upper panel; cyan histogram for blue star-forming,
red histogram for quiescent galaxies, both with and without
emission lines), the good separation between the colour distribution
of the different spectral classes, although with some overlap
for $2 < B-z < 3$. 
We verified whether the slit aperture
effects would bias the comparison between the 
observed colours and spectral classification: the
slit corrections, estimated on the basis of the difference
between the integrated magnitudes computed from the
flux-calibrated spectra and the I-band photometric data,
do not show any trend both with the magnitude and $B-z$ colour
in the analyzed sample.

The correlation between colour and spectral classification
becomes even tighter when we consider galaxy redshifts.
The $B-z$ \ versus redshift plot (lower panel; cyan and red squares
for star-forming and quiescent galaxies, respectively,
magenta crosses for intermediate ones) exhibits a clear segregation
between the galaxy classes. 
To divide effectively the two galaxy populations,
we adopted the $B-z$ \ versus redshift evolutionary track of a template
spectrum from the Coleman, Wu \& Weedman Extended set \citep{CWW,saw97}  
with properties corresponding most closely to those dividing the
two populations. The chosen template had the optical colours
\citep{fuk95} and spectral appearance \citep{ken92} of an Sab galaxy.
Also in local galaxy samples \citep[i.e. The Revised Shapley-Ames
Catalog of Galaxies;][]{vdb07}, the transition between blue and red
galaxies occurs at the Hubble type Sab/Sb.

The lower panel of Fig.~\ref{fig_Bz_sep} 
shows that almost all of the small number of discordant galaxies
(blue quiescent and red star-forming) fall close to the separation
track  in the colour classification plane, confirming the agreement
between the spectral and photometric classification.
Moreover, galaxies of the intermediate spectral class (red $D4000$
with emission lines, magenta crosses in Fig.~\ref{fig_Bz_sep})
have a colour distribution similar to that of quiescent galaxies,
in agreement with the observed similarity of the stellar continuum in
the composite spectra noted above (Fig.~\ref{fig_compo}). 

By merging these two spectral classes and using the track of
the Sab-template to separate the photometrically ``blue'' and
``red'' galaxies, we compiled a $2\times2$ contingency table 
for the high quality sample (Table~\ref{tbl-spepho}), 
useful in performing a statistical analysis of the relationship
between the spectral and photometric classification.
To quantify the degree of correlation between the two
classification schemes, we computed Pearson's linear correlation
coefficient for the data presented in Table~\ref{tbl-spepho}:
the two classifications are well correlated, as indicated
by a Pearson coefficient of 0.77.

\begin{table}[ht]
\caption{Numbers of galaxies in the spectroscopic and photometric classes}
\label{tbl-spepho}
\centering
\begin{tabular}{|c|ccc|}
\hline\hline
& & & \\[-6pt]& \multicolumn{3}{c|}{Spectral classification} \\
$B-z$ & Quiescent & Star-forming & TOTAL\\
\cline{2-4}
& & & \\[-6pt]
Red   & 119 & ~20 & 139 \\
Blue  & ~27 & 335 & 362 \\
TOTAL & 146 & 355 & 501 \\
\hline
\end{tabular}
\end{table}

In the following, we summarize how the different galaxy populations
inhabit the colour versus spectral classification plane:
\begin{itemize}
\item
82\% of objects classified as ``quiescent'' from  spectra are red;
\item
94\% of objects classified as ``star-forming'' from  spectra are blue;
\item
86\% of objects with red colours are ``quiescent'' from the spectra;
\item
93\% of objects with blue colours are ``star-forming'' from the spectra.
\item
The fraction of outliers is $\sim$9\% (47/501).
\end{itemize}
The correspondence between the two classifications
is not optimal for the 18\% quiescent
galaxies that show bluish colours. We investigated
the nature of these objects by studying their composite spectrum.
The continuum in the composite of blue quiescent galaxies
is largely indistinguishable from that of the red, quiescent galaxies,
but the spectrum shows a moderate \oii \ emission line ($EW_0 = 4.4$\AA)
and a strong $H\delta$ in absorption ($EW_0 = -3.8$\AA). 
The presence of these two spectral features represents star
formation activity, which could also explain the bluish colours
of this class of galaxy. 
The rest-frame equivalent width of \oii, measured in the composite
spectrum, is below the cut adopted for the spectral classification
($EW_0({\tt [OII]}) > 5$\AA), and only one of the 27 blue quiescent
galaxies is spectroscopically classified as intermediate. 
Therefore, it is more reasonable  to consider these objects as being
galaxies in-between the two main spectral classes, rather than galaxies
of extremely discordant spectral and colour properties.
Indeed, about 30\% of these galaxies
can be spectroscopically classified as post-starburst galaxies
and constitute the 64\% of the $1k$~sample of post-starburst
galaxies (7/11; Vergani et~al., in preparation).

The good agreement between the classification from
spectral data (quiescent/star-forming galaxies) and that based
on colours (red/blue galaxies) is achieved within the relatively
bright ($I_{AB} < 22.5$) zCOSMOS sample;
using the deeper ($I_{AB} < 24.0$) VVDS survey, \citet{fra07}
demonstrated that the incidence of red star-forming
or blue quiescent galaxies appear to increase when studying
fainter galaxy samples.

\section{ACS Morphology of galaxy spectral classes}

The HST/ACS imaging is an integral component of the COSMOS survey,
providing very high sensitivity and high resolution 
imaging over a large field of view. 
These space-based images yield resolved morphologies for
several hundreds of thousands galaxies, and the zCOSMOS-bright survey
magnitude limit ($I_{AB} < 22.5$) is matched with the I-band
ACS depth to allow classical bulge-disk decomposition.
%both via visual inspection and with computer-aided classification
%techniques. 
The COSMOS survey is a natural ``battlefield'' for competing
morphological classification schemes.
%for all the groups interested in the morphological analysis 
%of this huge galaxy sample. 
Among all the possible and equally good techniques implemented by 
the various teams within the COSMOS
collaboration, we selected as our morphological classification tool
the {\it Zurich Estimator of Structural Types} \citep[ZEST;][]{sca07a}.
We adopted this approach mainly because $ZEST$ classifies galaxy
types by applying a variety of non-parametric techniques
that measure physical parameters directly on the ACS images in a way
similar to how we classify galaxies in terms of both spectroscopic
and optical colour data.

In brief, $ZEST$ quantitatively describes the galaxy structure
using three variables (PC1; PC2; PC3), obtained by performing a
principal component analysis (PCA) in the five-dimensional
parameter space defined by asymmetry (A), concentration (C),
Gini coefficient (G), the 2nd-order moment of the brightest
20\% of galaxy pixels (M20), and the ellipticity of the light
distribution ($\epsilon$). 
%and the Sersic index ($n$)
%which parameterizes the two-dimensional galaxy surface
%brightness of the galaxy.
The morphological classification is performed in the
PC1-PC2-PC3 space by linking a (dominant) morphological class
to different regions in the PC-space.
Finally, $ZEST$ assigns each analyzed object to
a straightforward morphological Type (= 1 for early-type galaxies,
= 2 for disk galaxies and = 3 for irregular galaxies).
Furthermore, a ``bulgeness'' parameter (from 0 for bulge-dominated
galaxies to 3 for bulge-less disks) is also assigned to each
galaxy classified as Type 2 \citep{sca07a},
following the median value of the Sersic indices distribution
of galaxies clustered in the PC-space. The Sersic indices were
computed by \citet{sar07}.

%
% Figure 7
%
\begin{figure}
\resizebox{\hsize}{!}{\includegraphics{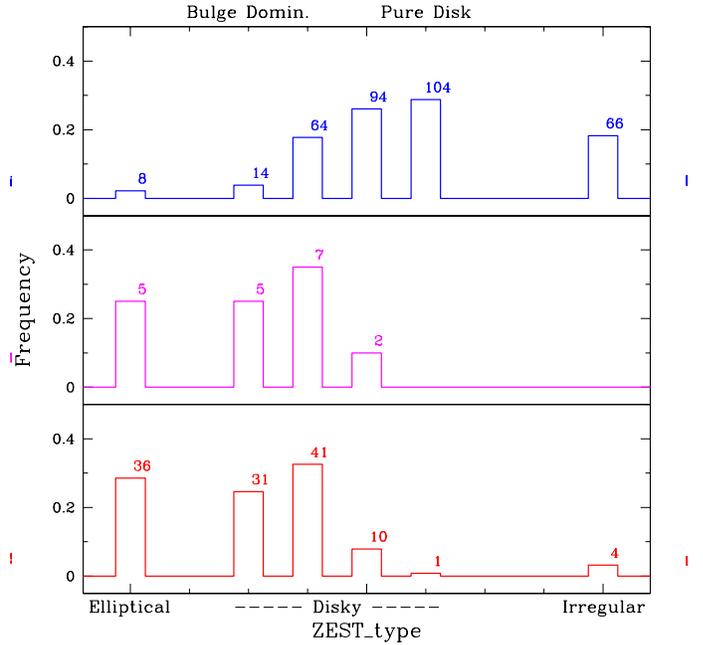}}
\caption{Frequency distributions of $ZEST$ morphological types 
for the spectroscopic blue emission line galaxies (top histogram),
the spectroscopic intermediate galaxies (i.e. galaxies which
show red continuum with emission lines; middle histogram),
and the quiescent galaxies (bottom histogram).
The four central bins represent the most-populated
$ZEST$ class of disk galaxies, further splitted on the basis of the
``bulgeness'' parameter. 
The numbers on top of the bars indicate the number of galaxies
in each bin.
\label{fig_spec_morph}}
\end{figure}

We compare the spectral classification with the morphological
characterization assigned by $ZEST$ in Fig.~\ref{fig_spec_morph},
where the histograms represent the frequency distributions of the
morphological types for the three main spectroscopic classes.
This figure illustrates a fairly good correlation between
the spectral and the morphological classification, 
represented by a gradual increase in the fraction of star-forming galaxies
towards morphological late-types, and, similarly, an increase in
spectroscopic {quiescent} galaxies towards the morphological early-types.
%
% Figure 8
%
\begin{figure}
\centering
\resizebox{\hsize}{!}{\includegraphics{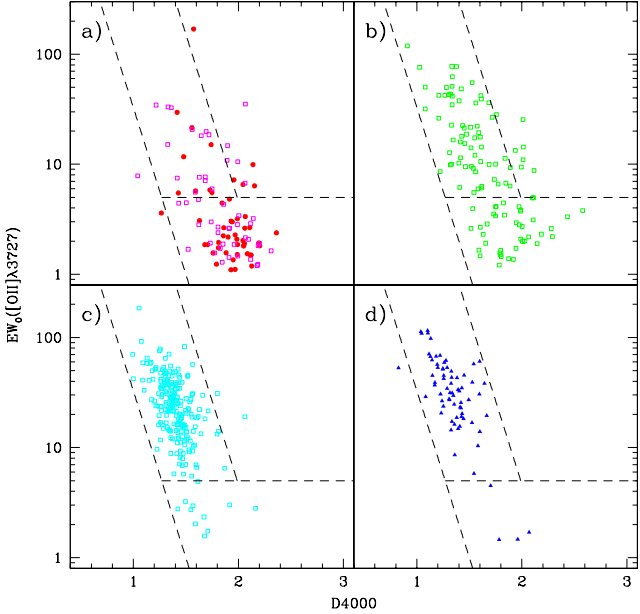}}
\caption{Distributions of different morphological groups in our
spectral classification plane. Panel a): Spheroidal galaxies,
including both pure ellipticals ($ZEST$-type~$=1$; filled circles)
and bulge-dominated galaxies ($ZEST$-type~$=2.0$; empty squares).
Panel b): disky galaxies with intermediate bulgeness
parameter ($ZEST$-type~$=2.1$; empty squares).
Panel c): ``disk-dominated'' galaxies ($ZEST$-type~$=2.2$ and $=2.3$;
empty squares).
Panel d): Irregular galaxies ($ZEST$-type~$=3$; filled triangles).
In all the four panels, the dashed lines indicate the spectral
class boundaries.
\label{fig_D4_O2_ZEST}}
\end{figure}

A more instructive way to look at the correspondence
between the morphological and spectral classifications is
presented in Fig.~\ref{fig_D4_O2_ZEST}, which shows the
distribution of the morphological types in our
spectral classification plane. 
Following the morphological characterization, we divided the galaxies into:
$(i\,)$~``spheroids'', a class including both 
the pure ellipticals ($ZEST$-type~$=1$) and the bulge-dominated
galaxies ($ZEST$-type~$=2$ with bulgeness parameter~$=0$)\footnote{
~the inclusion of the bulge-dominated disk galaxies into the same 
class as pure ellipticals was already proposed by \citet{sca07b}};
$(ii\,)$~``morphologically-mixed'' objects, disky 
($ZEST$-type~$=2$) galaxies with intermediate bulgeness
parameter ($=1$);
$(iii\,)$~``disk-dominated'' galaxies with $ZEST$-type~$=2$ and 
bulgeness parameter~$\ge2$;
and $(iv\,)$~``irregulars'', with $ZEST$-type~$=3$.
The four panels of Fig.~\ref{fig_D4_O2_ZEST} 
demonstrate that all but one of these morphologically
defined galaxy types appear to occupy the regions 
expected based on their spectral classification. 
% 36/39 (92%) Pure Ellipticals are quiescent from the spectra.
% 68/83 (82%) Spheroids are quiescent from the spectra.
% 173/185 (93%) Disk-dominated are starforming from the spectra.
% 61/64 (95%) Irregulars are starforming from the spectra.
More than 90\% of the morphological ellipticals are quiescent
from the spectra, as well as a large fraction (82\%) of the
spheroids. Also, more than 90\% of the disk-dominated and
irregular galaxies have blue emission line spectra. 
Only the  morphologically mixed objects (i.e. disk galaxies
with intermediate B/T parameters) are spread all over the
spectral classification plane.
Table~\ref{tbl-spemorph1} shows the distribution of $ZEST$ types
in the $1k$-galaxy sample as a function of their spectroscopic class.

\begin{table}[ht]
\caption{Numbers of galaxies in the spectroscopic and ZEST classes}
\centering
\label{tbl-spemorph1}
\begin{tabular}{|c|ccc|}
\hline\hline
& & & \\[-6pt]& \multicolumn{3}{c|}{Spectral classification} \\
$ZEST$-type & Quiescent & Star-forming & TOTAL\\
\cline{2-4}
& & & \\[-6pt]
1.0-2.0 &  76 & ~24 & 100 \\
2.1     &  55 & ~70 & 125 \\
2.2-3.0 & ~11 & 256 & 267 \\
TOTAL   & 142 & 350 & 492 \\
\hline
\end{tabular}
%\tablecomments{The total number of galaxies is different with
%respect to that in Table~\ref{tbl-spepho} because for some objects
%the morphological classification is not available.}
\end{table}

% Including the morphological class Type = 2.1 into the spheroidal 
% galaxies we obtain the following 2x2 contingency table: 
%    Spectral classification
%   Morphol.      Early   Starforming
%   Spheroids      116           65
%   Disk/Irr        15          234
%
% Including the morphological class Type = 2.1 into the disk/Irr galaxies 
% we obtain the following 2x2 contingency table: 
%    Spectral classification
%   Morphol.      Early   Starforming
%   Spheroids       68           15
%   Disk/Irr        63          284

The application of a $2\times2$ contingency table, as a statistical test
of correspondence between the spectral and morphological classifications,
is not a straightforward exercise in this case, since it is difficult
to assign objects with $ZEST$-type~$=2.1$ to an early- or late- galaxy type.
 
\begin{figure}
\resizebox{\hsize}{!}{\includegraphics{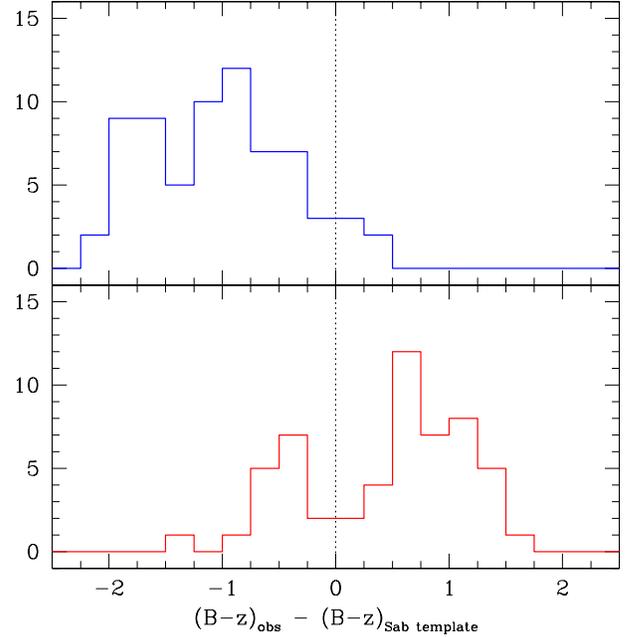}}
\caption{The distribution of the differences between the observed
$(B-Z)$ colour of the Type~$=2.1$ galaxies and the colour,
at the redshift of each galaxy, of the Sab template adopted to
photometrically separate the blue and red populations
in Fig.~\ref{fig_Bz_sep}. In the upper panel, the galaxies spectroscopically
classified as star-forming; in the lower panel, the 
galaxies spectroscopically classified as quiescent.
\label{fig_Bz_ZEST_2.1}}
\end{figure}

\Citet{bai08} discovered a similar population of morphologically
intermediate SDSS galaxies, finding that inclination effects play
an important role in their classification. Nevertheless, both a visual
inspection of the ACS snapshots and a statistical analysis of the 
asymmetry and ellipticity parameters do not suggest the presence of
an inclination bias in the zCOSMOS galaxies with $ZEST$-type~$=2.1$.
This intermediate morphological class is plausibly composed of galaxies
spanning a wide range of bulge-to-disk ratios and $ZEST$ alone is unable
to identify a dominant component in their visual morphology. 
We decided to fully utilize the classification cube potentialities, 
including the photometric colours, in the analysis of this intermediate
morphological class. Figure~\ref{fig_Bz_ZEST_2.1} shows
the colour distribution of $ZEST$-type~$=2.1$ galaxies, divided following the 
spectral classification. The upper panel shows the distribution
of the differences between the observed $(B-z)$ colour and the colour,
at the same redshift of each galaxy, of the Sab template
adopted to separate the blue and red populations in the
previous section, for galaxies spectroscopically classified as star-forming;
the lower panel shows the same colour difference for the 
Type~$=2.1$ galaxies spectroscopically classified as quiescent.
The figure clearly shows that galaxies with $ZEST$-type~$=2.1$
have $B-z$ colours largely in agreement with their spectral classification:
%red  early Type 2.1 = 39/55 70.90908813
%blue starf Type 2.1 = 65/70 92.85713959
93\% (65/70) of the $ZEST$ Type~$=2.1$ classified as ``star-forming''
from the spectra have blue colours, whereas 71\% (40/55) of the
$ZEST$ Type~$=2.1$ classified as ``quiescent'' from the spectra
have red colours. Interestingly, the latter galaxy class exhibits a
bimodal colour distribution, which is separated well by 
the colour track of the Sab~template. We investigated the average
properties of the two classes of red and blue quiescent galaxies
with intermediate ZEST morphology by means of the analysis of their composite
spectra: the stellar continuum appears similar in the two composites,
resembling that of the normal quiescent galaxies
(see Fig.~\ref{fig_compo}), even if the average spectrum of the blue
quiescent class is  of low signal-to-noise ratio because of the
small number of coadded galaxy spectra. The main
difference is the presence of emission lines in the blue quiescent
composite (with EW(\oii)$\sim$5\AA), while emission lines are undetected
in the red quiescent composite. However, the careful analysis of 
single galaxy sub-classes is beyond the scope of this paper, and 
%will be addressed in a forthcoming paper where the properties of such
%galaxy classes will be explored with a larger zCOSMOS data set.
are the subject of a forthcoming paper in which the properties of these
galaxy classes will be revisited using a larger zCOSMOS data set.
%
% Figure 9
%

\begin{table}[ht]
\caption{Numbers of galaxies in the spectroscopic and morphological classes}
\centering
\label{tbl-spemorph2}
\begin{tabular}{|c|ccc|}
\hline\hline
& & & \\[-6pt]& \multicolumn{3}{c|}{Spectral classification} \\
Morph+($B-z$) & Quiescent & Star-forming & TOTAL\\
\cline{2-4}
& & & \\[-6pt]Spheroids & 115 & ~29 & 144 \\
Disk/Irr  & ~27 & 321 & 348 \\
TOTAL     & 142 & 350 & 492 \\
\hline
\end{tabular}
%\tablecomments{The total number of galaxies is different with
%respect to that in Table~\ref{tbl-spepho} because for some objects
%the morphological classification is not available.}
\end{table}

We therefore separated the galaxies of this morphological class by colour, 
by assigning the red-coloured Type 2.1 galaxies to a Spheroid class
and the blue-coloured Type 2.1 galaxies to a Disk/Irregulars category.
Following this recipe, we derived a $2\times2$ contingency table 
(Table~\ref{tbl-spemorph2}), which provided a clear indication of the
correlation between the two classifications 
(Pearson correlation coefficient= 0.73).

\section{The galaxy classification cube}

It is evident from the previous sections that strong correlations
are present between the spectral classification and both the colour and 
morphological properties of the zCOSMOS-bright galaxies.
In this section, we merge these classifications
into a three-dimensional characterization, the galaxy classification cube.
We include in our analysis all galaxies with 
\hbox{$0.45\le z \le1.25$}~\footnote{since the
features adopted for spectral classification ($D4000$ and EW(\oii))
can be seen in the zCOSMOS-bright spectra only within this redshift range}
and complete photometric and morphological information. 
The full sample includes 577 galaxies, of which 487 (84\%) 
belong to the high-quality spectroscopic sample on the basis of the
assigned redshift flag.
   
A three-digit code was then assigned to each galaxy of the sample,
such that each digit represented one of the three classifications
(spectral, photometric, and morphological, respectively). 
Each digit can assume a bitwise value. This corresponds to 1, 
if the galaxy was spectroscopically classified as quiescent,
had red $B-z$ colour, or its morphology was dominated by the bulge component.
Alternatively, we assigned a digit value of 2
if the galaxy was spectroscopically star-forming, had blue
$B-z$ colour or showed a disky or irregular morphology.

We recall two caveats related to both the procedure we adopted
to separate galaxies into one of the two classes, and the presence
of galaxies that deviate from the bimodal
trends in properties (quiescent/star-forming, red/blue,
bulge-dominated/disk-dominated). 
First, a galaxy spectrum with emission lines but a red stellar
continuum (implied by a relatively high value of $D4000$) was
classified spectroscopically as quiescent because their global
properties indicated that they were probably ellipticals
experiencing a modest star-formation episode or including
a faint/obscured AGN. Secondly, we used the $B-z$ colour to
determine a more robust separation of the galaxies included
by ZEST in the intermediate class ($ZEST$~Type~$=2.1$). 

% Count_CCube
% ===========
% TOTAL 1302
% galaxies with spe-class   511 (631)
% galaxies with spe-class=0   4 (29)
% galaxies with spe-class without colour information   7 (9)
% galaxies with spe-class without morph information  13 (16)
% galaxies with full spec-phot-morph classification 487 (577)
% =================== Good objects
% galaxies 2-2-2  311 ( 63.9%)
% galaxies 1-1-1  105 ( 21.6%)
% galaxies 1-2-2   18 (  3.7%)
% galaxies 2-1-1   13 (  2.7%)
% galaxies 2-1-2    7 (  1.4%)
% galaxies 1-2-1    9 (  1.8%)
% galaxies 2-2-1   15 (  3.1%)
% galaxies 1-1-2    9 (  1.8%)
% ===================
% 487 galaxies 100.0%
% ===================
% =================== ALL objects
% galaxies 2-2-2  374 ( 64.8%)
% galaxies 1-1-1  113 ( 19.6%)
% galaxies 1-2-2   26 (  4.5%)
% galaxies 2-1-1   17 (  2.9%)
% galaxies 2-1-2    8 (  1.4%)
% galaxies 1-2-1   11 (  1.9%)
% galaxies 2-2-1   17 (  2.9%)
% galaxies 1-1-2   11 (  1.9%)
% ===================
% 577 galaxies 100.0%
% ===================

\begin{table}[ht]
\caption{The Classification Cube: galaxy counts and class percentages}
\centering
\label{tbl-ccube}
\begin{tabular}{|c|rc|rc|}
\hline\hline
& & & \\[-6pt]class-code & \multicolumn{2}{c|}{High-Q. Sample} &  \multicolumn{2}{c|}{~~~All Galaxies~~~} \\
 s-c-m & N~ & (\%) & N~ & (\%) \\
%\cline{2-4}
\hline
& & & & \\[-6pt]
2-2-2 & ~311 & (63.9) & ~374 & (64.8) \\
1-1-1 & ~105 & (21.6) & ~113 & (19.6) \\
1-2-2 &  ~18 &  (3.7) &  ~26 &  (4.5) \\
2-1-1 &  ~13 &  (2.7) &  ~17 &  (2.9) \\
2-1-2 &   ~7 &  (1.4) &   ~8 &  (1.4) \\
1-2-1 &   ~9 &  (1.8) &  ~11 &  (1.9) \\
2-2-1 &  ~15 &  (3.1) &  ~17 &  (2.9) \\
1-1-2 &   ~9 &  (1.8) &  ~11 &  (1.9) \\
\hline
& & & \\[-6pt]TOTAL & ~487 &  (100) & ~577 &  (100) \\
\hline
\end{tabular}
\end{table}

In Table~\ref{tbl-ccube}, numbers and percentages of objects 
belonging to each cell of the classification cube are presented,
using the three-digit code to identify the galaxy groups. The
numbers for both the high-quality and the full sample are listed,
and although all the single classification definitions % and divisions
were determined by using the high-quality sample, the percentages
of galaxies included in each group vary only slightly between the
two different samples. The main result emerging from the 3D analysis of
the optical properties of the zCOSMOS-bright galaxies is that a large
fraction of them shows a fully concordant classification: 85.5\%
(84.4\%) of the analyzed galaxies in the high-quality (full) 
sample can be assigned consistently to the red or blue branch
of the bimodal distribution using both a spectral, photometric,
and morphological criterion. Due to the redshift constraints
imposed by the spectral classification, we cannot explore a
very large redshift domain. However, we divided the useful
redshift range into two bins, so that each bin contained a
similar number of objects:
in the low-z range \hbox{($0.45\le z \le0.73$)} 83.4\% of the
galaxies shows a fully concordant classification, while in
the high-redshift range \hbox{($0.73\le z \le1.25$)} this
percentage slightly grows to 85.5\%. 
We conclude that the fraction of galaxies for which the three
classification schemes are in agreement remains constant
(within our statistical errors) for the entire redshift range
covered by our analysis.

\section{Conclusion}

We have presented a classification scheme for galaxies 
observed by the $1k$~zCOSMOS-bright sample, developed using spectral,
colour, and morphological information. Since the
spectral classification that we adopted is based on measurements
of the \oii \ and $D4000$ features, and given the wavelength range
of the zCOSMOS spectra, we applied this classification scheme to
galaxies in the redshift range \hbox{$0.45 \le z \le 1.25$}. 
We demonstrated that the adopted spectral classification
%being simple and straightforward to apply, 
is an effective way of separating galaxies into two main classes, 
due to the high quality of the zCOSMOS data.

We have found that an excellent correlation exists between the
classification from spectral data (quiescent/star-forming galaxies)
and that based on colours (red/blue galaxies), and that a similarly
good correlation exists between the classification
based on spectral data and that based on morphological analysis
(early-/late-type galaxies) only if we complement
the morphological classification with additional colour information.
The ``outliers'' in this simple 2-way analysis of different
classification methods (i.e. spectra versus colours and spectra
versus morphology) are quiescent galaxies (high $D4000$ value)
with emission lines (about 4\% of the total sample of
galaxies) and morphologically intermediate galaxies (about 20\% of the
total sample of galaxies) that consists of both blue/star-forming and
red/quiescent galaxies. These are the morphologically-classified
galaxies that require additional colour information to ensure good
agreement between the classification of schemes using both spectroscopic
and morphological data.

Finally, analyzing the distribution of all galaxies in %the sample in
a 3D-classification cube, based on the three different classification
methods, we have found that for about 85\% of the galaxies 
there is complete agreement between the classifications
(i.e. quiescent, red, ellipticals or bulge-dominated galaxies
($\sim$~20\%) or star-forming, blue, disk-dominated galaxies ($\sim$~65\%)). 
If additional colour information was not used for the $ZEST$-type~$=2.1$
galaxies (i.e. by assigning all of them to the
spheroid- or to the disk-galaxies), the percentage of fully concordant
classifications would decrease to about 75\%.

%
% Figure 10
%
\begin{figure}
\resizebox{\hsize}{0.9\hsize}{\includegraphics{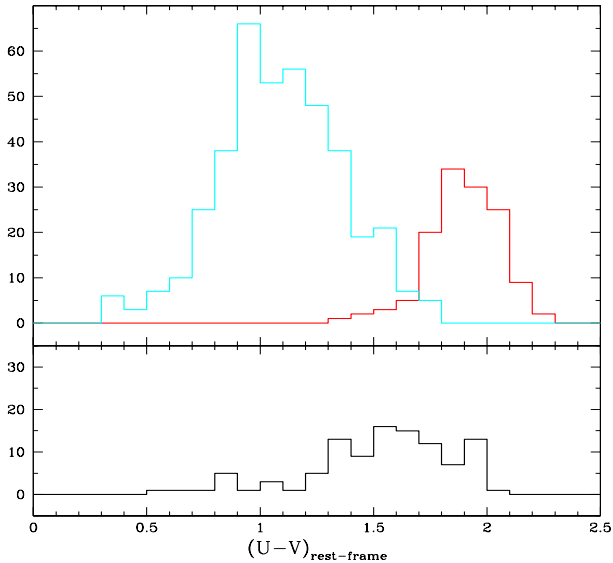}}
%\resizebox{\hsize}{!}{\includegraphics{hbimodal.eps}}
\caption{Classification cube and rest-frame $U-V$ colour of the $1k$~zCOSMOS-bright
data set. The upper panel shows the rest-frame colour distributions
of galaxies with fully concordant classification, the blue histogram
representing the star-forming, blue, disk-dominated galaxies,
whilst the red histogram is for the quiescent, red, ellipticals or
bulge-dominated galaxies. The lower panel shows the rest-frame colour
distribution for the outliers, i.e. galaxies which do not present the
same behaviour in all the three classifications.
\label{fig_hbimodal}}
\end{figure}

The correspondence between the classification cube and classical
representation of galaxy bimodality, which uses the rest-frame
$U-V$ colours\footnote{the rest-frame colours were computed
following the method developed by \citet{ilb05}}, is shown in
Fig.~\ref{fig_hbimodal}. In the upper panel, as expected, the 
rest-frame $U-V$ colour distributions of the two galaxy groups
with fully concordant classifications are separated well and reproduce
the bimodality. Perhaps more interesting is the distribution
of outliers in the lower panel, which is centered in-between the
two branches of the galaxy bimodality, in the so-called `green valley',
the relatively sparse region between the blue cloud of star-forming
galaxies, and the red sequence of quiescent ones.

The results presented in this paper demonstrate the potential of
the data set provided by the COSMOS/zCOSMOS survey for investigating 
the nature of the observed relationships between galaxy properties
and highlight fundamental aspects of the galaxy population. 
The main conclusion of this work is that the bimodality is a
prevalent feature of the galaxy population, at least out to $z\sim 1$,
in three critical parameters describing galaxy physical properties:
morphological, colour, and spectroscopic diagnostic line data.
More detailed analysis of the properties of
both a wider sample of galaxies and outliers in this
3D-classification cube, based on an analysis of the larger 10K
zCOSMOS-bright sample, will be presented elsewhere.

\begin{acknowledgements}

%We are grateful to the zCOSMOS team for the patience.
This work was partially supported by INAF under PRIN-2006/1.06.10.08
and by ASI under grant ASI/COFIS I/016/07/0.

\end{acknowledgements}

%
% The following command ends your manuscript. LaTeX will ignore any text
% that appears after it.

\end{document}